\documentclass[aps,pre,twocolumn,superscriptaddress,showpacs]{revtex4}
\usepackage{amsfonts,amssymb,amsmath,latexsym,epsfig,times} 
\begin{document}

[Phys. Rev. E {\bf 71}, 016116 (2005)]

\title{Network Synchronization, Diffusion, and the Paradox of Heterogeneity}

\author{Adilson E. Motter}
\email{motter@mpipks-dresden.mpg.de}
\affiliation{Max Planck Institute for the Physics of Complex Systems,
N\"othnitzer Strasse 38, 01187 Dresden, Germany}

\author{Changsong Zhou}
\email{cszhou@agnld.uni-potsdam.de}
\affiliation{Institute of Physics, University of Potsdam
PF 601553, 14415 Potsdam,  Germany}

\author{J\"{u}rgen Kurths}
\affiliation{Institute of Physics, University of Potsdam
PF 601553, 14415 Potsdam,  Germany}


\begin{abstract}

Many complex networks display strong heterogeneity in the degree (connectivity)
distribution.  Heterogeneity in the degree distribution often reduces the average
distance between nodes but, paradoxically, may suppress synchronization in
networks of oscillators coupled symmetrically with uniform coupling strength.
Here we offer a solution to this apparent paradox.  Our analysis is partially
 based on the
identification of a diffusive process underlying the communication between
oscillators and reveals a striking relation between this process and the
condition for the linear stability of the synchronized states.  We show that,
for a given degree distribution, the maximum synchronizability is achieved when
the network of couplings is weighted and directed, and the overall cost involved
in the couplings is minimum.  This enhanced synchronizability is solely
determined by the mean degree and does not depend on the degree distribution and
system size.  Numerical verification of the main results is provided for
representative classes of small-world and scale-free networks.

\end{abstract}


\maketitle

\section{Introduction}

The interplay between network structure and dynamics has attracted a great deal
of attention in connection with a variety of processes \cite{rev}, including
epidemic spreading \cite{epd}, congestion and cascading failures \cite{casc},
and synchronization of coupled oscillators
\cite{watts:book,sync,BP:2002,NMLH:2003,GDLP:2000,Wu:2003,heter,ROH:2004}.  Much
of this interest has been prompted by the discovery that numerous real-world
networks \cite{rev} share universal structural features, such as the small-world
\cite{sw} and scale-free properties \cite{sf}.
Small-world networks
(SWNs) exhibit a small average distance between nodes and high clustering
\cite{sw}.  Scale-free networks (SFNs) are characterized by an algebraic, highly
heterogeneous distribution of degrees (number of links per node) \cite{sf}.
Most SFNs also exhibit a small average distance between nodes \cite{sw_sf} and
this distance
may
become smaller as the heterogeneity (variance) of the degree distribution is
increased \cite{CH:2003}.  It has been shown that these structural properties
strongly influence the dynamics on the network.

In oscillator networks, the ability to synchronize is generally enhanced in both
SWNs and random SFNs as compared to regular lattices \cite{com1}.  However, it
was recently shown that random  networks with strong heterogeneity in the degree
distribution, such as  random SFNs, are much more difficult to synchronize than random
homogeneous networks \cite{NMLH:2003}, even though the former display smaller
average distance between nodes \cite{CH:2003}.  This result is interesting for
two main reasons.  First, it challenges previous interpretations that the
enhancement of synchronizability in SWNs and SFNs would be due to the reduction
of the average distance between oscillators.  Second, in networks where
synchronization is desirable, it puts in check the hypothesis that the scale-free
property has been favored by evolution for being dynamically advantageous.

Previous work has focused mainly on the role played by shortest paths between
nodes.  By considering only shortest paths it is implicitly assumed that the
information spreads only along them.  However, the communication between
oscillators is more closely related to a process of diffusion on the network,
which is a process involving all possible paths between nodes.  Another basic
assumption of previous work is that the oscillators are coupled symmetrically
and with the same coupling strength \cite{NMLH:2003}.  Under the assumption of
symmetric coupling, the maximum synchronizability may be indeed achieved when the
coupling strength is uniform \cite{Wu:2003}.  But to get a better
synchronizability, the couplings are not necessarily symmetrical.  Many
real-world networks are actually directed \cite{rev} and weighed \cite{w}, and
the communication capacity of a node is likely to saturate when the degree
becomes large.

In this paper, we study the effect that asymmetry and saturation of coupling
strength have on the synchronizability of complex networks.  We identify a
physical process of information diffusion that is relevant for the communication
between oscillators and we investigate the relation between this process and the
stability of synchronized states in directed networks with weighted couplings.
We address these fundamental issues using as a paradigm the problem of complete
synchronization of identical oscillators.

We find that the synchronizability is explicitly related to the mixing rate of the
underlying diffusive process.  For a given degree distribution, the
synchronizability is maximum when the diffusion has a uniform stationary state,
which in general requires the network of couplings to be weighted and directed.
For large  sufficiently random networks, the maximum synchronizability is
primarily determined by the mean degree of the network and does not depend on
the degree distribution and system size, in sharp contrast with the case of
unweighted (symmetric) coupling, where the synchronizability is strongly suppressed as the
heterogeneity or number of oscillators is increased.  Furthermore, we show that
the total cost involved in the network coupling is significantly reduced, as
compared to the case of unweighted coupling, and is minimum when the
synchronizability is maximum. Some of these results were announced in
Ref.~\cite{preprint}.

The fact that the communication between oscillators takes place along all
paths explains why the synchronizability does not necessarily correlate with the
average distance between oscillators.  Moreover, the synchronizability of SFNs
is strongly enhanced when the network of couplings is suitably weighted.  This,
in addition to the well know improved structural robustness of SFNs
\cite{attack}, may have played a crucial role in the evolution of many SFNs.

The paper is organized as follows.
In Sec.~\ref{s2}, we introduce the
 synchronization model and the measure of synchronizability. 
In Sec.~\ref{s3}, we study the corresponding process of diffusion.
In Sec.~\ref{s4}, we focus on the case of maximum synchronizability.
The problem of cost is considered in Sec.~\ref{s5}.
In Sec.~\ref{s6}, we present direct simulations on networks of maps.
Discussion and conclusions are presented in the last section.

\section{Formulation of the Problem}
\label{s2}

We introduce a generic model of coupled oscillators and we present a condition
for the linear stability of the synchronized states in terms of the eigenvalues of
the coupling matrix.

\subsection{Synchronization Model}

We consider complete synchronization of linearly coupled identical oscillators:
\begin{equation}
\frac{dx_i}{dt}=f(x_i)-\sigma\sum_{j=1}^{N}G_{ij}h(x_j), \;\;\; i=1,2,\ldots N,
\label{eq1}
\end{equation}
where $f=f(x)$ describes the dynamics of each individual oscillator, $h=h(x)$ is
the output function, $G=(G_{ij})$ is the coupling matrix, and $\sigma$ is the
overall coupling strength.  The rows of matrix $G$ are assumed to
have zero sum to ensure that the synchronized state 
$\{ x_i(t)=s(t), \forall i \; | \; ds/dt=f(s) \}$
is a solution of Eq.~(\ref{eq1}).

In the case of symmetrically coupled oscillators with uniform coupling strength,
$G$ is the usual (symmetric) Laplacian matrix $L=(L_{ij})$:  the diagonal
entries are $L_{ii} = k_i$, where $k_i$ is the degree of node $i$, and the
off-diagonal entries are $L_{ij}=-1$ if nodes $i$ and $j$ are connected and
$L_{ij}=0$ otherwise.  For $G_{ij}=L_{ij}$, heterogeneity in
the degree distribution suppresses synchronization in important
classes of networks \cite{NMLH:2003} (see also
Ref.~\cite{heter}).  The synchronizability can be easily enhanced if we modify
the topology of the network of couplings.  Here, however, we address the problem
of enhancement of synchronizability for a {\it given} network topology.

In order to enhance the synchronizability of heterogeneous networks, we propose
to scale the coupling strength by a function of the degree of the nodes. 
For specificity, we take
\begin{equation}
G_{ij}=L_{ij}/k_i^{\beta},
\label{eq2}
\end{equation}
where $\beta$ is a tunable parameter.  We say that the network or coupling is
weighted when $\beta\neq 0$ and unweighted when $\beta=0$.  The underlying
network associated with the Laplacian matrix $L$ is undirected and unweighted,
but for $\beta\neq 0$, the network of couplings becomes not only weighted but
also directed because the resulting matrix $G$ is in general asymmetric. 
This is a special kind of directed network where the number of {\it
in}-links is equal to the number of {\it out}-links in each node, and the
directions are encoded in the strengths of in- and out-links.  In spite of 
the possible asymmetry of matrix $G$,
all the eigenvalues of matrix $G$ are nonnegative reals and can be ordered as
$0=\lambda_1\le \lambda_2\cdots \le\lambda_N$, as shown below.

\subsection{Basic Spectral Properties}

Eq.~(\ref{eq2}) can be written as
\begin{equation}
G=D^{-\beta}L,
\label{eq3}
\end{equation}
where $D=$ diag$\{k_1,k_2,\ldots k_N\}$ is the diagonal matrix of degrees. 
(We recall that the degree $k_i$ is the number of
oscillators coupled to oscillator $i$.) From the
identity $\det (D^{-\beta}L-\lambda I)=\det (D^{-\beta/2}LD^{-\beta/2}-\lambda
I)$, valid for any $\lambda$, where $\det$ denotes the determinant and $I$ is
the $N\times N$ identity matrix, we have that the spectrum of eigenvalues of
matrix $G$ is equal to the spectrum of a symmetric matrix defined as
\begin{equation}
H=D^{-\beta/2}LD^{-\beta/2}.
\label{eq4}
\end{equation}
That is,
$\rho(G)=\rho(H),$
where $\rho$ denotes the set of eigenvalues.  From this follows that all
eigenvalues of matrix $G$ are real, as anticipated above.  It is worth
mentioning that, although the eigenvalues of $G$ and $H$ are equal, from the
numerical point of view it is much more efficient to compute the eigenvalues
from the symmetric matrix $H$ than from $G$.

Additionally, all the eigenvalues of matrix $G$ are nonnegative because $H$ is
positive semidefinite, and the smallest eigenvalue $\lambda_1$ is always zero
because the rows of $G$ have zero sum.  Moreover, if the network is connected,
then $\lambda_2 > 0$ for any finite $\beta$.  This follows from the corresponding
property for $L$ and Eq.~ (\ref{eq4}), i.e., the fact that matrices $H$ and $L$
are congruent.  Naturally, the study of complete synchronization of the whole
network only makes sense if the network is connected.

For $\beta=1$, matrix $H$ is the normalized Laplacian matrix studied in spectral
graph theory \cite{chung:book}.  In this case, if $N \ge 2$ and the network is
connected, then $0<\lambda_2\le N/(N-1)$ and $ N/(N-1)\le\lambda_N\le 2$.  For
spectral properties of unweighted SFNs, see
Refs.~\cite{other_ref,CLV:2003,DGMS:2003}.

\subsection{Synchronizability}

The variational equations governing the linear stability of a synchronized state
$\{ x_i(t)=s(t), \forall i\}$ of the system in Eqs.~(\ref{eq1}) and (\ref{eq2})
can be diagonalized into $N$ blocks of the form
\begin{equation}
\frac{d\eta}{dt}=\left[ Df(s) - \alpha Dh(s)\right]\eta,
\label{eq6}
\end{equation}
where $D$ denotes the Jacobian matrix, $\alpha=\sigma\lambda_i$, and $\lambda_i$
are the eigenvalues of the coupling matrix $G$.  The largest Lyapunov exponent
$\Gamma(\alpha)$ of this equation can be regarded as a master stability
function, which determines the linear stability of the synchronized state for
any linear coupling scheme \cite{msf}:  the synchronized state is stable if
$\Gamma(\sigma\lambda_i)<0$ for $i=2,\ldots N$.  (The eigenvalue $\lambda_1$
corresponds to a mode parallel to the synchronization manifold.)  

For many widely studied oscillatory systems \cite{BP:2002,msf}, the master
stability function $\Gamma(\alpha)$ is negative in a single, finite interval
$(\alpha_1,\alpha_2)$.  Therefore, the network is synchronizable for some
$\sigma$ when the eigenratio
$R=\lambda_N/\lambda_2$
  satisfies
\begin{equation}
R <\alpha_2/\alpha_1.
\label{eq7}
\end{equation}
The right-hand side of this equation depends only on the dynamics ($f$, $h$, and
$s$), while the eigenratio $R$ depends only on the coupling matrix $G$.  The
problem of synchronization is then reduced to the analysis of eigenvalues of the
coupling matrix \cite{BP:2002}:  the smaller the eigenratio $R$, the larger the
synchronizability of the network, and vice versa.

\section{Diffusion and Balance of Heterogeneity}
\label{s3}

We study a process of diffusion relevant for the communication between
oscillators and we argue that the synchronizability is maximum ($R$ is minimum)
for $\beta=1$.

\subsection{Diffusion Process}
\label{s3a}

From the identity
\begin{equation}
{\sum_{j=1}^{N}G_{ij}h(x_j) = \sum_{j=1}^{N}k_i^{-\beta} A_{ij} [h(x_i)-h(x_j)]}
\label{eq10}
\end{equation}
we observe that the weighted coupling scheme in Eqs.~(\ref{eq1}) and
(\ref{eq2}) is naturally related to a diffusive process with absorption and
emission described by the transition matrix
\begin{equation}
P=\frac{1}{\Lambda}D^{-\beta}A,
\label{eq11}
\end{equation}
where $A=D-L$ is the adjacency matrix, and $\Lambda$ is the largest eigenvalue
of $D^{-\beta}A$.
According to this process, if we start with an arbitrary distribution
$y =(y^{(1)}, \ldots  y^{(N)})$, where $y^{(i)}$ is associated with the
initial state at node $i$, after $n$ time steps the distribution is $P^ny$.
This process is different from the usual (conservative) random walk process.  In
particular, because $\sum_i P_{ij}$ may be different from 1, the diffusion {\it
in} and {\it out} of a node may differ even in the stationary state.

For instance, consider a network of three nodes, $a$-$b$-$c$, where nodes $a$ and
$c$ have degree $1$ and are both connected to node $b$.  For $\beta=1$, the
transition matrix is
\begin{equation}
P = \left( \begin{array}{ccc}
	0 & 1 & 0   \\
      1/2 & 0 & 1/2 \\
        0 & 1 & 0   \\
           \end{array}
    \right).
\label{eq12}
\end{equation}
In the stationary state, each of the three nodes has, say, one unity (of the
``diffusive quantity").  At each time step, node $b$ receives $1/2$ unit from
node $a$ and $1/2$ unit from node $c$, and each of the nodes $a$ and $c$
receives $1$ unit from node $b$.  Therefore, node $b$ sends a total of $2$ units
and receives only $1$ unit, while each of the nodes $a$ and $c$ sends $1/2$ unit
and receives $1$ unit.  This means that there is an ``absorption" of $1/2$ unit
at each of the nodes $a$ and $c$ and the ``emission" of $1$ unit at node $b$.
It is in this sense that the matrix in Eq.~(\ref{eq11}) describes a diffusive
process with absorption and emission.

On the other hand, the usual random walk process is conservative at each node.
Such a process is described by the matrix $D^{-\beta}AC^{-1}$, where
$C_{ij}=\delta_{ij}\sum_{\ell\sim j}{k_{\ell}}^{-\beta}$ and the sum is over all
the $k_j$ nodes connected to node $j$.  For a node $i$ connected to a node $j$
and a uniform distribution, this conservation law implies that the amount of
information that node $i$ receives from node $j$ depends on the degree of all
the nodes connected to node $j$.  But this is not what happens in a network of
self-sustained oscillators.  In a network of oscillators, the amount of
information that oscillator $i$ receives directly from oscillator $j$ can only
depend on the strength of the coupling from $j$ to $i$, which is proportional to
$1/k_i^{\beta}$, as in the process described by the transition matrix in
Eq.~(\ref{eq11}).

\subsection{Balance of Heterogeneity}
\label{s3b}

Because the master stability function $\Gamma(\alpha)$ is negative in a finite
interval $(\alpha_1,\alpha_2)$, increasing (decreasing) the overall coupling
strength $\sigma$ beyond a critical value $\sigma_{max}$ ($\sigma_{min}$)
destabilizes the synchronized state.  Dynamically, the loss of stability is due
to a short (long) wavelength bifurcation at $\sigma=\sigma_{max}$ ($\sigma_{min}$)
\cite{short_long} (see also Ref.~\cite{ROH:2004}). 
Physically, this bifurcation excites the shortest (longest)
spatial wavelength mode because some oscillators are too strongly (weakly)
influenced by the others.

Now, consider the process of diffusion described by matrix $P$ on a network
where not all the nodes have the same degree.  Starting with an arbitrary
distribution $y$, after $n$ steps we have $P^ny$.  If we require the (stationary)
distribution for $n\rightarrow \infty$ to be uniform, we obtain $\beta=1$ because this is the only case
where $y_0=(1,1,\ldots 1)$ is an eigenvector associated with the eigenvalue $1$,
which is the largest eigenvalue of matrix $P$.  For $\beta<1$, the distribution
is more heavily concentrated on nodes with large degree. For $\beta>1$, the
concentration happens on nodes with small degree.  Physically, this means that
for both $\beta<1$ and $\beta>1$ some oscillators are more strongly influenced
than others and the ability of the network to synchronize is limited by the
least and most influenced oscillators:  for small (large) $\sigma$ the system is
expected to undergo a long (short) wavelength bifurcation due to the least
(most) influenced nodes, as explained above.  We then expect the network to
achieve maximum synchronizability at $\beta=1$.

\begin{figure}[pt]
\begin{center}
\epsfig{figure=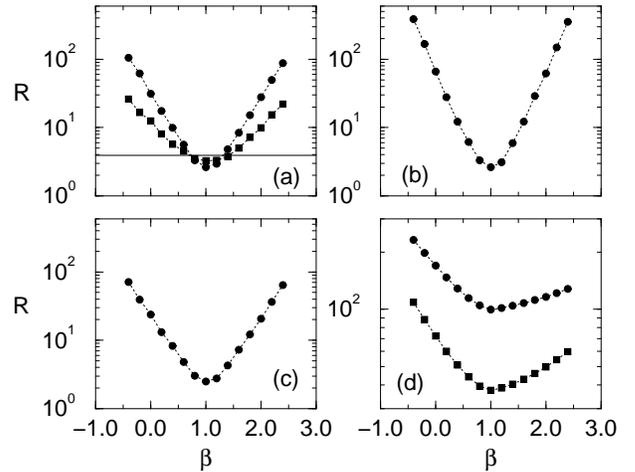,width=8.0cm}
\caption{
Eigenratio $R$ 
as a function of $\beta$:  
(a) random SFNs with $\gamma=3$
($\bullet$), $\gamma=5$ (${\scriptscriptstyle \blacksquare}$),
and $\gamma=\infty$ (solid line), for $k_{min}=10$; 
(b) networks with expected
scale-free sequence 
for $\gamma=3$ and ${\tilde
k}_{min}=10$;
(c) growing SFNs
for $\gamma=3$ and $m=10$;
(d)  SWNs with $M=256$ ($\bullet$) and $M=512$ (${\scriptscriptstyle \blacksquare}$), for $\kappa=1$.
Each curve is the result of an average over 50 realizations for $N=1024$.
}
\label{fig1}
\end{center}
\end{figure}

In Fig.~\ref{fig1} we show the numerical verification of this hypothesis for
various models of complex networks.  The networks are built as follows
\cite{connected}:

\begin{description}
\item {({\it i})} 
{\it Random SFNs} \cite{NSW:2001} --- Each node is assigned to have a number
$k_i \ge k_{min}$ of ``half-links'' according to the probability distribution $P(k)\sim
k^{-\gamma}$, where $\gamma$ is a scaling exponent and $k_{min}$ is a constant integer.  The
network is generated by randomly connecting these half-links to form links,
prohibiting self- and repeated links.  In the limit $\gamma=\infty$, all
nodes have the same degree $k= k_{min}$.

\item {({\it ii})}
{\it Networks with expected scale-free sequence } \cite{CLV:2003} --- The
network is generated from a sequence $\tilde{k}_1, \tilde{k}_2, \ldots
\tilde{k}_N$, where $\tilde{k}_i\ge \tilde{k}_{min}$ follows the distribution
$P(\tilde{k})\sim \tilde{k}^{-\gamma}$ and $\max_i\tilde{k}_i^2<\sum_i
\tilde{k}_i$.
A link is then independently assigned to each pair of nodes
$(i,j)$  with probability $p_{ij}=\tilde{k}_i\tilde{k}_j/\sum_i \tilde{k}_i$.
In
this model, self-links are allowed.  We observe, however, that the eigenratio
$R$ is insensitive to the removal of self-links.

\item {({\it iii})}
{\it Growing SFNs} \cite{LLYD:2002} --- We start with a fully
connected network with $m$ nodes and at each time step a new node with $m$ links
is added to the network.  Each new link is connected to a node $i$ in the
network with probability $\Pi_i\sim (1-p) \hat{k}_i + p$, where $\hat{k}_i$ is
the updated degree of node $i$ and $0\le p \le 1$ is a tunable parameter.  For
large degrees, the scaling exponent of the resulting network is
$\gamma=3+p[m(1-p)]^{-1}$.  For $p=0$, the exponent is $\gamma=3$ and we recover the
Barab\'asi-Albert model \cite{sf}.


\item {({\it iv})}
{\it SWNs} \cite{NMW:2000} ---
Starting with a ring of $N$ nodes, where each node is connected to $2\kappa$
first neighbors, we add $M\le N(N-2\kappa -1)/2$ new links between randomly
chosen pairs of nodes.  Self- and repeated links are avoided.  \end{description}

Our extensive numerical computation on the models ({\it i-iv}) shows that the
eigenratio $R$ has a well defined minimum at $\beta=1$ in each case
[Fig.~\ref{fig1}].  The only exception is the class of homogeneous networks,
where all the nodes have the same degree $k$.  When the network is homogeneous,
the weights $k_i^{-\beta}$ can be factored out in Eq.~(\ref{eq11}) and a uniform
stationary distribution is achieved for any $\beta$.  In this case, the
eigenratio $R$ is independent of $\beta$, as shown in Fig.~\ref{fig1}(a) for
random homogeneous networks with $k=10$ (solid line).  A random homogeneous
network corresponds to a random SFN for $\gamma=\infty$.  In all other cases,
including the relatively homogeneous SWNs, the eigenratio exhibits a pronounced
minimum at $\beta=1$ (note the logarithmic scale in Fig.~\ref{fig1}).

In SWNs, the heterogeneity of the degree distribution increases as the number
$M$ of random links is increased.  The eigenratio $R$ at $\beta=0$ reduces as
$M$ is increased, but the eigenratio at $\beta=1$ reduces even more, so that the minimum of
the eigenratio becomes more pronounced as the heterogeneity of the degree
distribution is increased [Fig.~\ref{fig1}(d)].
Similar results are observed in the original Watts-Strogatz model of
SWNs, where the mean degree is kept fixed as the number
of random links is increased  \cite{sw}. SWNs of pulse oscillators also
present enhanced synchronization at $\beta=1$ \cite{GDLP:2000}.

In SFNs, the heterogeneity increases as the scaling exponent $\gamma$ is
reduced.  As shown in Fig.~\ref{fig1}(a) for random SFNs, the minimum of the
eigenratio $R$ becomes more pronounced as the heterogeneity of the degree
distribution is increased.  The same tendency is observed across different
models of networks.  For example, for a given $\gamma$ and
$k_{min}=\tilde{k}_{min}$, the minimum of the eigenratio is more pronounced in
networks with expected scale-free sequence [Fig.~\ref{fig1}(b)] than in random
SFNs [Fig.~\ref{fig1}(a)], because the former may have nodes with degree smaller
than $\tilde{k}_{min}$.  For small $\gamma$, the eigenratio in growing SFNs
[Fig.~\ref{fig1}(c)] behaves similarly to the eigenratio in random SFNs
[Fig.~\ref{fig1}(a)].  A pronounced minimum for the eigenratio $R$ at $\beta=1$
is also observed in various other models of complex networks \cite{in_out}.

\subsection{Mean Field Approximation}

A mean field approximation provides further insight into the effects of degree
heterogeneity and the dependence of $R$ on $\beta$.

The dynamical equations (\ref{eq1}) can be rewritten as:
\begin{equation}
\frac{dx_i}{dt}=f(x_i)+\sigma k_i^{1-\beta} [\bar {h}_i-h(x_i)],
\label{eq13}
\end{equation}
where 
\begin{equation} 
\bar {h}_i=\frac{1}{k_i} \sum_j A_{ij} h(x_j)
\label{eq14}
\end{equation}
is the local mean field from all the nearest neighbors of oscillator $i$.
If the network is sufficiently random and the system is close to the
synchronized state $s$, we may assume that $\bar {h}_i\approx h(s)$ and we may
approximate Eq.~(\ref{eq13}) as
\begin{equation}
\frac{dx_i}{dt}=f(x_i)+\sigma k_i^{1-\beta}[h(s)-h(x_i)],
\label{eq15}
\end{equation}
indicating that the oscillators are decoupled and forced by 
a common  oscillator with output $h(s)$. 

From a variational equation analogous to Eq.~(\ref{eq6}), we have that all
oscillators in Eq.~(\ref{eq15}) will be synchronized by the common forcing when
\begin{equation} 
\alpha_1<\sigma k_i^{1-\beta}<\alpha_2 \;\;\; \forall i.
\label{eq16}
\end{equation}
For $\beta\ne 1$, it is enough to have a single node with degree very different
from the others for this condition not to be satisfied for any $\sigma$.  In
this case, the complete synchronization becomes impossible because the
corresponding oscillator cannot be synchronized.  Within this approximation, the
eigenratio is $R=(k_{max}/k_{min})^{1-\beta}$ for $\beta\leq 1$ and
$R=(k_{min}/k_{max})^{1-\beta}$ for $\beta > 1$, where $k_{min}=\min_i \{ k_i
\}$ and $k_{max}=\max_i \{ k_i \}$.  The minimum of $R$ is indeed achieved at
$\beta=1$, in agreement with our numerical simulations.

Therefore, this simple mean field approximation not only explains the results of
Ref.~\cite{NMLH:2003} on the suppression of synchronizability due to
heterogeneity in unweighted networks, but also predicts the correct condition
for maximum synchronizability in weighted networks.

\section{Mixing Rate and Synchronizability}
\label{s4}

We relate the eingenratio $R$ to the mixing rate of the process of diffusion
introduced in Sec.~\ref{s3a} and we argue that, for $\beta=1$ and large,
sufficiently random networks, the synchronizability depends only on the mean
degree of the network.

\subsection{Mean Degree Approximation}
\label{s4a}

We now present a general physical theory for the eigenratio $R$.  In what
follows we focus on the case of maximum synchronizability ($\beta=1$).
For $\beta=1$ and an arbitrary network,
Eq.~(\ref{eq11}) can be written as
\begin{equation}
P=D^{-1/2}(I-H)D^{1/2},
\label{eq20}
\end{equation}
where $H=D^{-1/2}LD^{-1/2}$ as in Eq.~(\ref{eq4}).
From the identity $\det (P-\lambda I)=\det (I-H-\lambda I)$, valid for any
$\lambda$, it follows that the spectra of matrices $P$ and $H$ are related via
$\rho(P)=1-\rho(H)$,
and the uniform stationary state of the process $P^n y$ is
associated with the null eigenvalue of the coupling matrix $G$.
We then define the mixing rate as $\nu=\ln\mu^{-1}$, where
\begin{equation}
\mu=\lim_{n\mapsto\infty}\|P^ny-y_0\|^{1/n}
\label{eq21}
\end{equation}
is the mixing parameter, $y_0=(1,1,\ldots 1)$ is the stationary distribution
discussed in Sec.~\ref{s3b}, $\; y$ is an arbitrary initial distribution
normalized as $\sum_i k_i y_i=\sum_i k_i $, and $\|\cdot \|$ is the usual
Euclidean norm.  In non-bipartite connected
networks \cite{chung:book},  we have $\lambda_N<2$ and the initial distribution $y$ always
converges to the stationary distribution $y_0$.

The convergence of the limit in Eq.~(\ref{eq21}) is dominated by the second
largest eigenvalue of matrix $P$ in absolute value, namely $\max_{i=2,\ldots
N}|1- \lambda_i|$. (The largest eigenvalue is associated with the stationary state $y_0$.)
As a result, for {\it any} network, the mixing parameter is
\begin{equation}
\mu= \max \{1-\lambda_2,\lambda_N-1\}.
\label{eq22}
\end{equation}
Therefore, the mixing is faster in networks where the eigenvalues of the
coupling matrix are concentrated close to $1$. 

The condition for the stability of the synchronized states also requires the
eigenvalues of the coupling matrix to be close to $1$, although through a slightly
different relation
($R=\lambda_N/\lambda_2$ to be small).  We can combine these two conditions to
write an upper bound for the eigenratio $R$ in terms of the mixing parameter:
\begin{equation}
R \le \frac{1+\mu}{1-\mu}.
\label{eq23}
\end{equation}
This relation is relevant because of its general validity and clear physical
interpretation.  We show that this upper bound is a very good approximation of
the actual value of $R$ in many networks of interest.

The mixing parameter $\mu$ can be expressed as an explicit function of the mean
degree $k$. Based on results of Ref.~\cite{CLV:2003} for random networks with
given expected degrees, we get
\begin{equation}
\max \{1-\lambda_2,\lambda_N-1\}= [1+o(1)]\frac{2}{\sqrt{k}}.
\label{eq24}
\end{equation}
Moreover, the semicircle law holds and the spectrum of matrix $P$ is symmetric
around $1$ for $k_{min}\gg \sqrt{k}$ in the thermodynamical limit \cite{CLV:2003}.
These results are rigorous for ensembles of networks with a given expected
degree sequence and sufficiently large minimum degree $k_{min}$, but our extensive numerical
computation supports the hypothesis
that the approximate relations
\begin{equation}
\lambda_2 \approx 1 - \frac{2}{\sqrt{k}}, \;\;\; 
\lambda_N \approx 1 + \frac{2}{\sqrt{k}}.
\label{eq25}
\end{equation}
hold under much milder conditions.  In particular, relations (\ref{eq25}) are
expected to hold true for any large, sufficiently random network with $k_{min}\gg 1$.  The rationale
for this is that, for $\beta=1$, the diffusion {\it in} each node of one such network
is the same as in a random homogeneous network with the same mean degree,
where relations (\ref{eq25}) are known to be satisfied \cite{chung:book}.

Under the assumption that $1-\lambda_2\approx \lambda_N-1$, the eigenratio
can be written as
\begin{equation}
R \approx \frac{1+\mu}{1-\mu},
\label{eq26}
\end{equation}
where $\mu$ is defined in Eq.~(\ref{eq22}).
Therefore, the larger the mixing rate (smaller $\mu$), the more synchronizable
the network (smaller $R$), and vice versa.  From Eq.~(\ref{eq25}), we have that
the mixing parameter can be approximated as $\mu \approx
2/\sqrt{k}$ and the eigenratio can be approximated as
\begin{equation}
R \approx \frac{1+2/\sqrt{k}}{1-2/\sqrt{k}}.
\label{eq27}
\end{equation}
Therefore, for $\beta=1$, the eigenratio $R$ is primarily determined by the mean
degree and does not depend on the number of oscillators and the details of the
degree distribution.

This is a remarkable result because, regardless of the degree distribution, the
network at $\beta=1$ is just as synchronizable as a random homogeneous network
with the same mean degree, and random homogeneous networks appear to be
asymptotically optimal in the sense that $R$ approaches the absolute
lower bound in the thermodynamical limit for large enough $k$ \cite{Wu:2003}.

\subsection{Numerical Verification}

Now we test our predictions in the models ({\it i-iii}) of SFNs, and we show that
the synchronizability is significantly enhanced 
for $\beta=1$ as compared to the case of unweighted coupling
($\beta=0$).

\begin{figure}[pt]
\begin{center}
\epsfig{figure=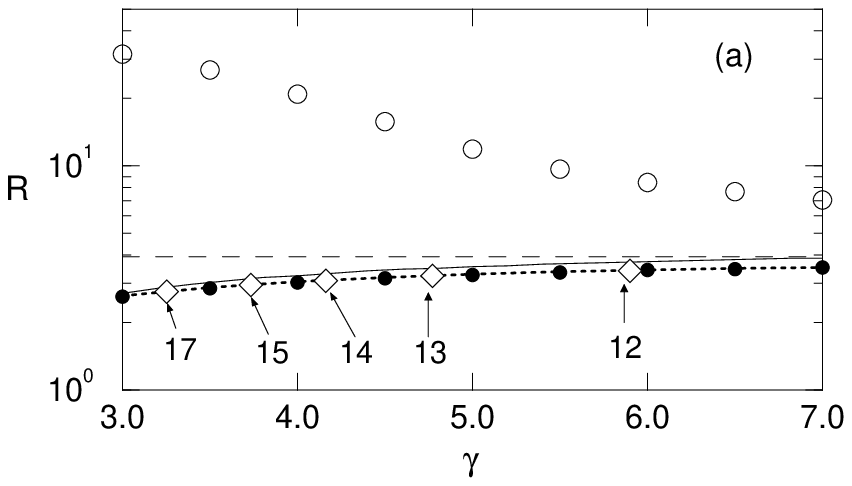,width=6.0cm}
\epsfig{figure=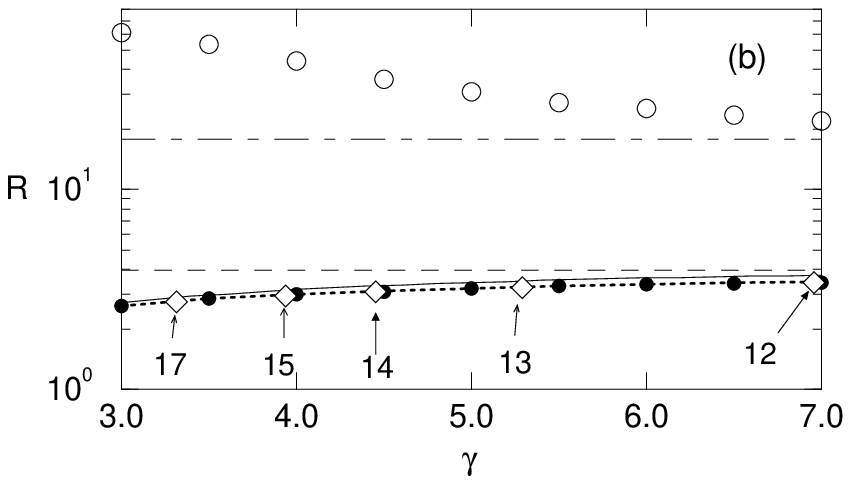,width=6.0cm}
\epsfig{figure=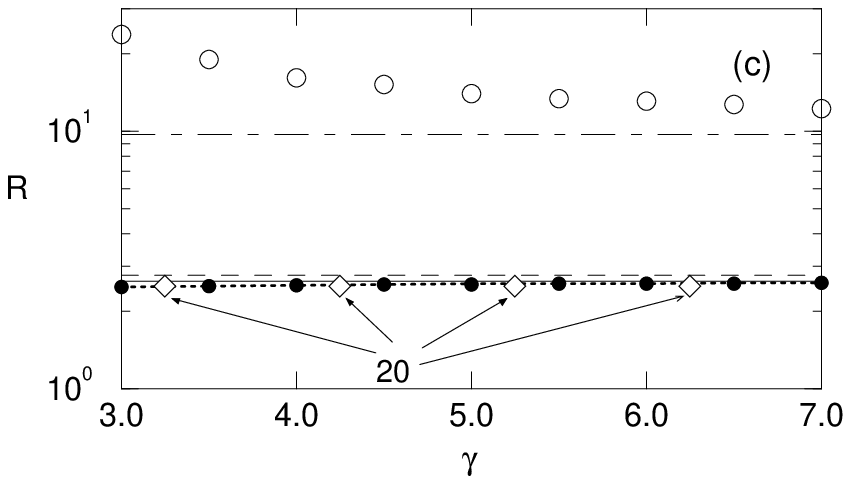,width=6.0cm}
\caption{
 Eigenratio $R$ as a function of the scaling exponent $\gamma$:  (a) random
 SFNs, (b) networks with expected scale-free sequence, and (c) growing SFNs, for
 $\beta=1$ ($\bullet$) and $\beta=0\;$ ($\circ$). The other curves are
 the approximations of the eigenratio in Eqs.~(\ref{eq26}) (dotted lines) and (\ref{eq27}) (solid lines),
 and the eigenratio  for $\beta=1$ (dashed lines) and
 $\beta=0$ (dot-dashed lines) at $\gamma=\infty$.
 The ${\scriptstyle \lozenge}$ symbols correspond
 to random homogeneous networks with the same mean degree of the
 corresponding SFNs (the degrees are indicated in the figure).
 The other network parameters are the same as in Fig.~\ref{fig1}.
}
\label{fig2}
\end{center}
\end{figure}

As shown in Fig.~\ref{fig2}, in unweighted SFNs, the eigenratio $R$ increases
with increasing heterogeneity of the degree distribution (see also
Ref.~\cite{NMLH:2003}).  But, as shown in the same figure, the eigenratio does
not increase
with heterogeneity when the coupling is
weighted at $\beta=1$.  The difference is especially large for small scaling
exponent $\gamma$, where the variance of the degree distribution is large and
the network is highly heterogeneous (note that Fig.~\ref{fig2} is plotted in
logarithmic scale).  The network becomes more homogeneous as $\gamma$ is
increased.  In the limit $\gamma=\infty$, random SFNs converge to random
homogeneous networks with the same degree $k_{min}$ for all the nodes
[Fig.~\ref{fig2}(a)], while networks with expected scale-free sequence converge
to Erd{\H o}s-R\'enyi random networks \cite{ballobas:book}, which have links
assigned with the same probability between each pair of nodes
[Fig.~\ref{fig2}(b)], and growing SFNs converge to growing random networks,
which are growing networks with uniform random attachment [Fig.~\ref{fig2}(c)].
As one can see from Figs.~\ref{fig2}(b) and \ref{fig2}(c), the synchronizability is
strongly enhanced even in the relatively homogeneous Erd{\H o}s-R\'enyi and
growing random networks; such an enhancement occurs also in SWNs.

For $\beta=1$, the eigenratio $R$ is well approximated by the relations in
Eq.~(\ref{eq26}) [Fig.~\ref{fig2}, dotted lines] and Eq.~(\ref{eq27})
[Fig.~\ref{fig2}, solid lines] for all three models of SFNs.  This confirms our
result that the synchronizability is strongly related to the mixing properties of
the network \cite{anybeta}.  For $\beta=1$, the eigenratio of the SFNs is also
very well approximated by the eigenratio of random homogeneous networks with the
same number of links [Fig.~\ref{fig2}, ${\scriptstyle \lozenge}$].  Therefore,
for $\beta=1$, the variation of the eigenratio $R$ with the heterogeneity of the
degree distribution in SFNs is mainly due to the variation of the mean degree of
the networks, which increases in both random SFNs and networks with expected
scale-free sequence as the scaling exponent $\gamma$ is reduced
[Figs.~\ref{fig2}(a) and \ref{fig2}(b)].

\begin{figure}[pt]
\begin{center}
\epsfig{figure=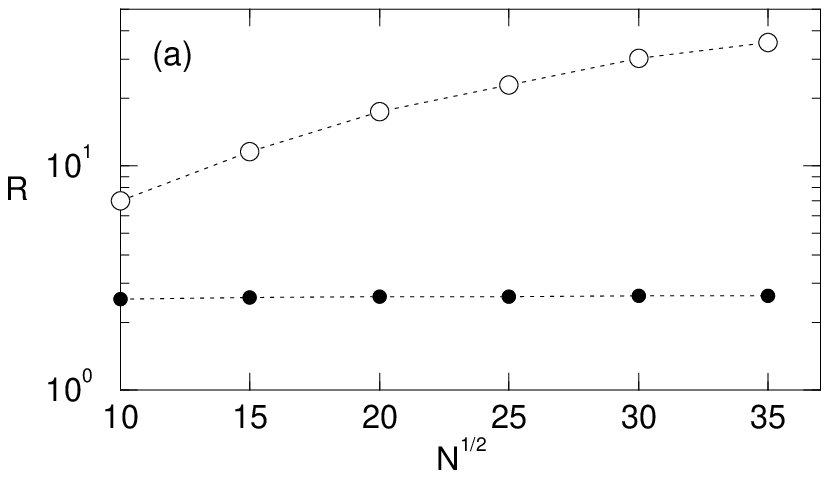,width=6cm}
\epsfig{figure=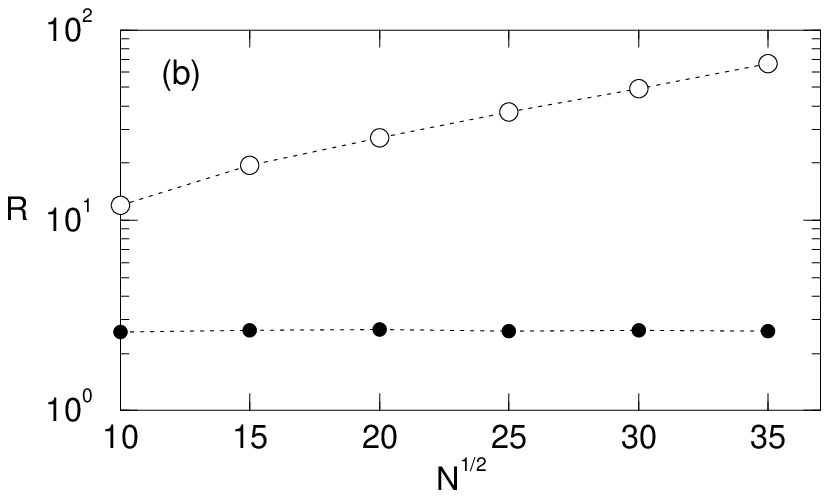,width=6cm}
\epsfig{figure=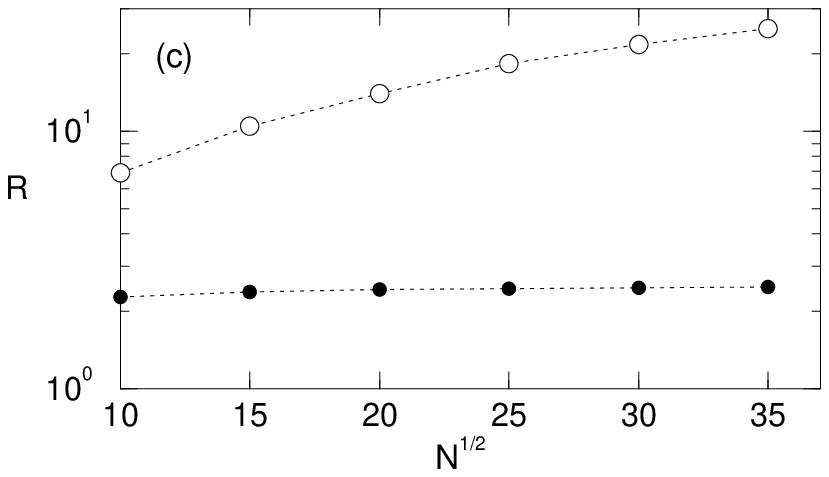,width=6cm}
\caption{Eigenratio $R$
 as a function of the number of oscillators for $\gamma=3$ and the SFN models in Figs.~\ref{fig2}(a)-(c),
 respectively. Dotted lines are guides for the eyes.
 The legend and other parameters are the same as in Fig.~\ref{fig2}.
}
\label{fig2.5}
\end{center}
\end{figure}

In Fig.~\ref{fig2.5}, we show the eigenratio $R$ as a function of the system
size $N$.  In unweighted SFNs, the eigenratio increases strongly as the number
of oscillators is increased.  Therefore, it may be very difficult or even
impossible to synchronize large unweighted networks.  However, for $\beta=1$,
the eigenratio of large networks appears to be independent of the system size,
as shown in Fig.~\ref{fig2.5} for the models ({\it i-iii}) of SFNs.  Similar
results are observed in many other models of complex networks.  All together,
these provide strong evidence for our theory.

\subsection{General Bounds}

We present bounds valid for {\it any} network weighted at $\beta=1$.  In this
case, if the network is connected but not globally connected and $N>2$, we have
\begin{equation}
1+ (N-1)^{-1}\le R\le 2NkD_{max},
\label{eq28}
\end{equation}
where $k$ is the mean degree and $D_{max}$ is the diameter of the network
(maximum distance between nodes).  This relation follows from the bounds $1/
NkD_{max}\le \lambda_2 \le 1$ and $1 + (N-1)^{-1} \le \lambda_N \le 2$
\cite{chung:book}.  For being valid for any network regardless of its structure,
the bounds in Eq.~(\ref{eq28}) are not tight for specific network models, such
as random homogeneous networks.  Nevertheless, they provide some insight into the problem.
In particular, heterogeneity in the degree distribution is not disadvantageous
in this case because it generally reduces the upper bound in Eq.~(\ref{eq28}),
and this is a major difference from the case of unweighted networks considered
previously \cite{NMLH:2003}.

\section{Coupling Cost}
\label{s5}

Having shown that weighted networks exhibit improved synchronizability, we now
turn to the problem of cost.  We show that the total cost involved in the
network of couplings is minimum at the point of maximum synchronizability
($\beta=1$).

The total cost $C$ involved in the network of couplings is defined as
the minimum (in the synchronization region) of the total strength of all
directed links,
\begin{equation}
 C=\sigma_{min}\sum_{i=1}^{N}k_i^{1-\beta},
\label{eq30}
\end{equation}
where $\sigma_{min}=\alpha_1/\lambda_2$ is the minimum coupling strength
for the network to synchronize.  We recall that $\alpha_1$ is the point
where the master stability function first becomes negative.  
For $\beta=1$, we have $C=N\alpha_1/\lambda_2$. 

\begin{figure}[pt]
\begin{center}
\epsfig{figure=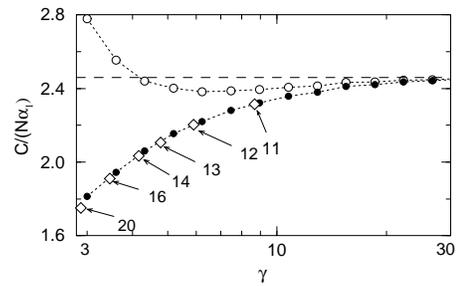,width=6.0cm}
\caption{Normalized cost $C/N\alpha_1$ as a function of the scaling exponent $\gamma$ for
random SFNs with $\beta=1$ ($\bullet$) and $\beta=0$ ($\circ$),
and for random homogeneous networks with the same mean degree (${\scriptstyle
\lozenge}$).  The 
dashed line corresponds to $\gamma=\infty$.
The other parameters are the same as in Fig.~\ref{fig1}.
}
\label{fig4}
\end{center}
\end{figure}

In heterogeneous networks, the cost at $\beta=1$ is significantly reduced as
compared to the case of unweighted coupling ($\beta=0$), as shown in
Fig.~\ref{fig4} for random SFNs.  The difference becomes more pronounced when
the scaling exponent $\gamma$ is reduced and the degree distribution becomes
more heterogeneous.  The cost for SFNs at $\beta=1$ is very well approximated by
the cost for random homogeneous networks with the same mean degree
[Fig.~\ref{fig4}, ${\scriptstyle \lozenge}$], in agreement with our analysis in
Sec.~\ref{s4a} that,
at $\beta=1$, the eigenvalue $\lambda_2$ is fairly independent of the degree distribution. 

\begin{figure}[pt]
\begin{center}
\epsfig{figure=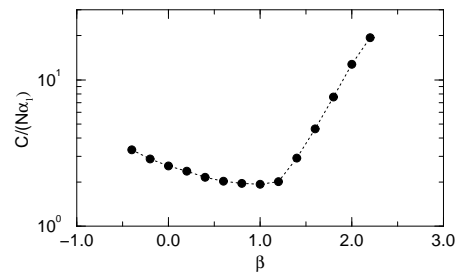,width=6.0cm}
\caption{Normalized cost $C/N\alpha_1$ as a function of $\beta$
for random SFNs with scaling exponent $\gamma=3$. The other parameters
are the same as in Fig.~\ref{fig1}.
}
\label{fig5}
\end{center}
\end{figure}

As a function of $\beta$, the cost 
has a 
broad
minimum
at $\beta=1$, as
shown in Fig.~\ref{fig5} for random SFNs.  Similar result is observed in other
models of complex networks, including the models ({\it ii-iv}) introduced in
Sec.~\ref{s3b}.  This result is important because it shows that maximum
synchronizability and minimum cost occur exactly at the same point.  Therefore,
cost reduction is another important advantage of suitably weighted networks.

\section{Direct Simulations}
\label{s6}

To confirm our analysis of enhanced synchronizability, we simulate
the dynamics on networks of chaotic maps.

The example we consider consists of SFNs of logistic maps,
$x_{n+1}=f(x_n)=ax_n(1-x_n)$, where the output function is
taken to be $h(x)=f(x)$.  In this case, the master
stability function is negative for $(1-e^{-\Gamma_0})=\alpha_1 <\alpha
<\alpha_2=(1+e^{-\Gamma_0})$, where $\Gamma_0>0$ is the Lyapunov exponent of the
isolated chaotic map.  In the simulations of the dynamics, the maps are assigned
to have random initial conditions close to the synchronization manifold.

We consider two values of the bifurcation parameter $a$ for which the logistic
map is chaotic:  $a=3.58$, where $\alpha_2/\alpha_1\approx 19$, and $a=4.0$,\
where $\alpha_2/\alpha_1=3$.  In both cases, our simulations show that, if
$R<\alpha_2/\alpha_1$, then there is a finite interval of the overall coupling
strength $\sigma_{min} <\sigma <\sigma_{max}$ where the network becomes
completely synchronized after a transient time.  Moreover, the simulations
confirm that $\sigma_{min}=\alpha_1/\lambda_2$ and $\sigma_{max}=
\alpha_2/\lambda_N$, as expected. In order to display the synchronization regions
for different $\beta$ in the same figure, we introduce
$\sigma^*=\sigma\sum_i k_i^{1-\beta}$, $\sigma_{min}^*=\sigma_{min}\sum_i
k_i^{1-\beta}$ and $\sigma_{max}^*=\sigma_{max}\sum_i k_i^{1-\beta}$.

\begin{figure}[pt]
\begin{center}
\epsfig{figure=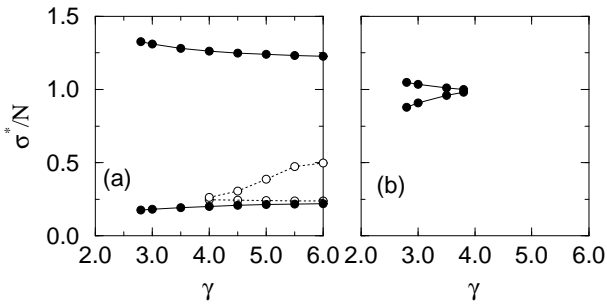,width=8.0cm}
\caption{Synchronization region ($\sigma_{min}^*, \sigma_{max}^*$) 
as a function of $\gamma>2.8$
in random SFNs of
logistic maps,
for $\beta=1$ ($\bullet$) and $\beta=0$ ($\circ$).
The bifurcation parameter is (a) $a=3.58$ and (b) $a=4.0$.
 Averages are taken over 20 realizations of the
networks.
The other parameters are the same as in Fig.~\ref{fig1}.
}
\label{fig6}
\end{center}
\end{figure}

In Fig.~\ref{fig6}, we show the synchronization region $(\sigma_{min}^*,
\sigma_{max}^*)$ as a function of the scaling exponent $\gamma$ in random SFNs,
for $\beta=1$ ($\bullet$) and $\beta=0$ ($\circ$). 
The factor $\sum_i k_i^{1-\beta}$ is $N$
for $\beta=1$ and $kN$ for $\beta=0$.
For $a=3.58$, the networks are
synchronizable for some $\sigma*$ in the region $\gamma\gtrsim 4.0$ if the
couplings are unweighted ($\beta=0$) and in a wider region of $\gamma$ if the
couplings are weighted at $\beta=1$ [Fig.~\ref{fig6}(a)].  In the region where
both weighted and unweighted networks are synchronizable, the interval
$(\sigma_{min}^*, \sigma_{max}^*)$, in which synchronization is
achieved, is much wider for $\beta=1$ than for $\beta=0$.  In terms of the
original coupling strength $\sigma$, the difference is even larger ($k$ times larger).
More strikingly, for $a=4.0$, unweighted networks do not synchronize for any
coupling strength, but the networks weighted at $\beta=1$ do synchronize for
$\gamma\lesssim 4.0$ in a nonzero interval of $\sigma^*$ [Fig.~\ref{fig6}(b)].

The cost $C$ defined in Eq.~(\ref{eq30}) is exactly $\sigma_{min}^*$.  In
agreement with the results in Fig.~\ref{fig4}, $\sigma_{min}^*$ is clearly
smaller for $\beta=1$ than for $\beta=0$ in the interval of $\gamma$ where both
weighted and unweighted networks are synchronizable [Fig.~\ref{fig6}(a)].  All
together, the results in Fig.~\ref{fig6} illustrate the enhancement of
synchronizability and the reduction of cost in weighted networks.

\section{Concluding Remarks}
\label{s7}

Motivated by the problem of complex-network synchronization, we have introduced
a model of directed networks with weighted couplings that incorporates the
saturation of connection strength expected in highly connected nodes of
realistic networks.  In this model, the total strength of all {\it in}-links at a node
$i$ with degree $k_i$ is proportional to $k_i^{1-\beta}$, where the parameter
$\beta$ is a measure of the degree-dependent saturation in the amount of
information that a node receives from other nodes.  In a network of oscillators,
the weights $k_i^{1-\beta}$ can be alternatively interpreted as a property of the
({\it input} function of the) oscillators rather than a property of the links.
We believe that this model can serve as a paradigm to address many problems of
dynamics on complex networks.

Here we have studied complete synchronization of identical oscillators.  We have
shown that, for a given network topology, the synchronizability is maximum and
the total cost involved in the network of couplings is minimum when $\beta=1$.
For large, sufficiently random network with minimum degree $k_{min}\gg 1$, the
synchronizability at $\beta=1$ is mainly determined by the mean degree and is
fairly independent of the number of oscillators and the details of the degree
distribution.

This should be contrasted with the case of unweighted coupling ($\beta=0$),
where the synchronizability is strongly suppressed as the number of oscillators
or heterogeneity of the degree distribution is increased.  In the case
$\beta=1$, the heterogeneity of the degree distribution is completely balanced
and the networks are just as synchronizable as random homogeneous networks with
the same mean degree. 

Our results are naturally interpreted within a framework where the condition for
the linear stability of synchronized states is related to the mixing rate of a
diffusive process relevant for the communication between oscillators.  
Under mild conditions, we have shown that, the larger the mixing
rate, the more synchronizable the network.  In particular, in unweighted
networks, the mixing rate decreases with increasing heterogeneity. 
This, along
with the condition $\beta=1$ for enhanced synchronizability, explains and solves
what we call the ``paradox of heterogeneity.'' This paradox refers to the
(apparently) paradoxal relation between the synchronizability and the average
distance between oscillators, observed in heterogeneous unweighted networks
\cite{NMLH:2003}, and is clarified when we observe that synchronizability is
ultimately related to the mixing properties of the network.

We expect our results to be relevant for both network design and the
understanding of dynamics in natural systems, such as neuronal networks, where
the saturation of connection strength is expected to be important.  Although we
have focoused mainly on SFNs, a class of networks that has received most
attention, our analysis is general and applies to networks with arbitrary degree
distribution.

\acknowledgements

The authors thank Takashi Nishikawa and Diego Paz\'o for valuable discussions
and for revising the manuscript.
A. E. M. was supported by Max-Planck-Institut f\"ur Physik komplexer Systeme.
C. S. Z. and J. K. were partially supported by SFB 555. 
C. S. Z. was also supported by the VW Foundation.


\begin{references}
\bibitem{anybeta} The mixing parameter $\mu$ can be defined for any $\beta$.
For arbitrary $\beta$, from an equation analogous to Eq.~(\ref{eq21}), we have
$\mu=|\Lambda'|/|\Lambda|$, where $\Lambda$ and $\Lambda'$ are (in absolute
value) the largest and second largest eigenvalues of matrix $D^{-\beta}A$.  Our
numerical results suggest that, for sufficiently large random SFNs, the mixing rate is
positively correlated with synchronizability for any $\beta$.  In particular,
the mixing rate decreases (increases) with decreasing $\gamma$ for $\beta=0$
($\beta=1$).  As a function of $\beta$, in the thermodynamical limit, the mixing
rate seems to converge to a maximum at $\beta=1$, in agreement with the enhanced
synchronizability observed in this case.
\end{references}
\end{document}